\documentclass{article}

\usepackage{PRIMEarxiv}
\usepackage{authblk}
\usepackage[utf8]{inputenc} 
\usepackage[T1]{fontenc}    
\usepackage{hyperref}       
\usepackage{url}            
\usepackage{booktabs}       
\usepackage{amsfonts}       
\usepackage{nicefrac}       
\usepackage{microtype}      
\usepackage{lipsum}
\usepackage{fancyhdr}       
\usepackage{graphicx}       
\graphicspath{{media/}}     

\usepackage{cite}

\usepackage{hyperref}
\hypersetup{
  colorlinks=true,
  linkcolor=black,
  citecolor=black,
  urlcolor=black
}
\title{Amyloid-Beta Axial Plane PET Synthesis from Structural MRI: An Image Translation Approach for Screening Alzheimer’s Disease
\thanks{2023 International Society of Magnetic Resonance in Medicine.  Toronto, Canada,  June 2-9.  Abstract Number 5040}
}

\author[1-4]{Fernando Vega }
\author[1-4]{Abdoljalil Addeh}
\author[1-4]{M. Ethan MacDonald}

\affil[1]{Department of Biomedical Engineering, Schulich School of Engineering, University of Calgary, Calgary, AB, Canada}
\affil[2]{Department of Electrical \& Software Engineering, Schulich School of Engineering, University of Calgary, Calgary, AB, Canada}
\affil[3]{Hotchkiss Brain Institute, Cumming School of Medicine, University of Calgary, Calgary, AB, Canada}
\affil[4]{Department of Radiology, Cumming School of Medicine, University of Calgary, Calgary, AB, Canada}

\begin{document}
\maketitle
\section*{Synopsis}
In this work, an image translation model is implemented to produce synthetic amyloid-beta PET images from structural MRI that are quantitatively accurate. Image pairs of amyloid-beta PET and structural MRI were used to train the model. We found that the synthetic PET images could be produced with a high degree of similarity to truth in terms of shape, contrast  and overall high SSIM and PSNR.  This work demonstrates that performing structural to quantitative image translation is feasible to enable the access of amyloid-beta information from only MRI.
\section*{Introduction}
Alzheimer’s Disease (AD) is the most prevalent cause of dementia, with an estimated economic burden of \$305 billion in 2020 in the US \cite{econo} . By the time that AD is clinically diagnosed, the patient is already experiencing neuronal loss and brain atrophy.   Amyloid-beta is a key molecule that is widely believed to be a root cause of AD pathophysiology that starts aggregating within the brain before clinical symptoms manifest \cite{abpathway, abstructure}.

\noindent Over the last 15 years, it has become possible to image amyloid-beta with the use of positron emission tomography (PET) tracers \cite{abpractice}, however, PET imaging has some disadvantages such as: cost (\$5000-to-\$8000 per scan), invasiveness, as it requires the injection of a radiotracer leading to radiation exposure, and the tracers are not available in many jurisdictions \cite{effectivedose};  these disadvantages limit the use for screening and early onset detection. 

\noindent Structural MRI is about 10-times cheaper than PET ($500-to-$800) and can aid to assess AD by examining structural changes and atrophy \cite{biomalz,spattialpatterns}. MRI has limited ability to provide molecular information, but image translation algorithms might be used to compute synthetic amyloid-beta PET images from structural MRI.  The translation is possible due to the relationships between amyloid-beta burden and brain atrophy \cite{relationab}. 

\noindent Image translation is an advanced form of machine learning that aims to render one image type from another, enabling the access to images that might not otherwise be captured \cite{image_trans}. These algorithms are commonly implemented using Convolutional Neural Networks (CNN) \cite{CNNrad} with Conditional Generative Adversarial Networks (cGAN) \cite{image-trans-CVPRs} as they excel in generating realistic images \cite{mygod}. cGANs require pairs of images that share mutual information \cite{GAN_review}. 

\noindent This work demonstrates a pipeline using cGAN to generate quantitatively accurate synthetic amyloid-beta PET images from structural MRI images. 

\section*{Method}
The Open Access of Imaging Studies (OASIS-3) \cite{OASIS} dataset provided 929  subjects with pairs of T1-weighted MRI and amyloid-beta PET images. From these, 609 are cognitively normal (CN) subjects and 489 at different stages of cognitive decline.  The MRI images were obtained with three different scanners: Siemens Biograph mMR 3T, Siemens Trio Tim 3T, and Siemens Sonata 1.5T with a resolution of 1mmx1mmx1mm for the 3T scanners and 1.2mmx1.2mmx1.2mm for the 1.5T scanners.   The amyloid-beta PET images were obtained with two scanners: Siemens Biograph 40 PET/CT and Siemens ECAT 962 using Pittsburgh Compound B \cite{PiB} as a radiotracer with an administered dose that ranged between 6 – 20 millicuries (mCI) and a 60-minute dynamic PET scan where Standard Uptake Value Ratio (SUVR) was obtained using the PET Unified Pipeline \cite{PUP}.

\begin{figure}[p]
\includegraphics[width=\textwidth]{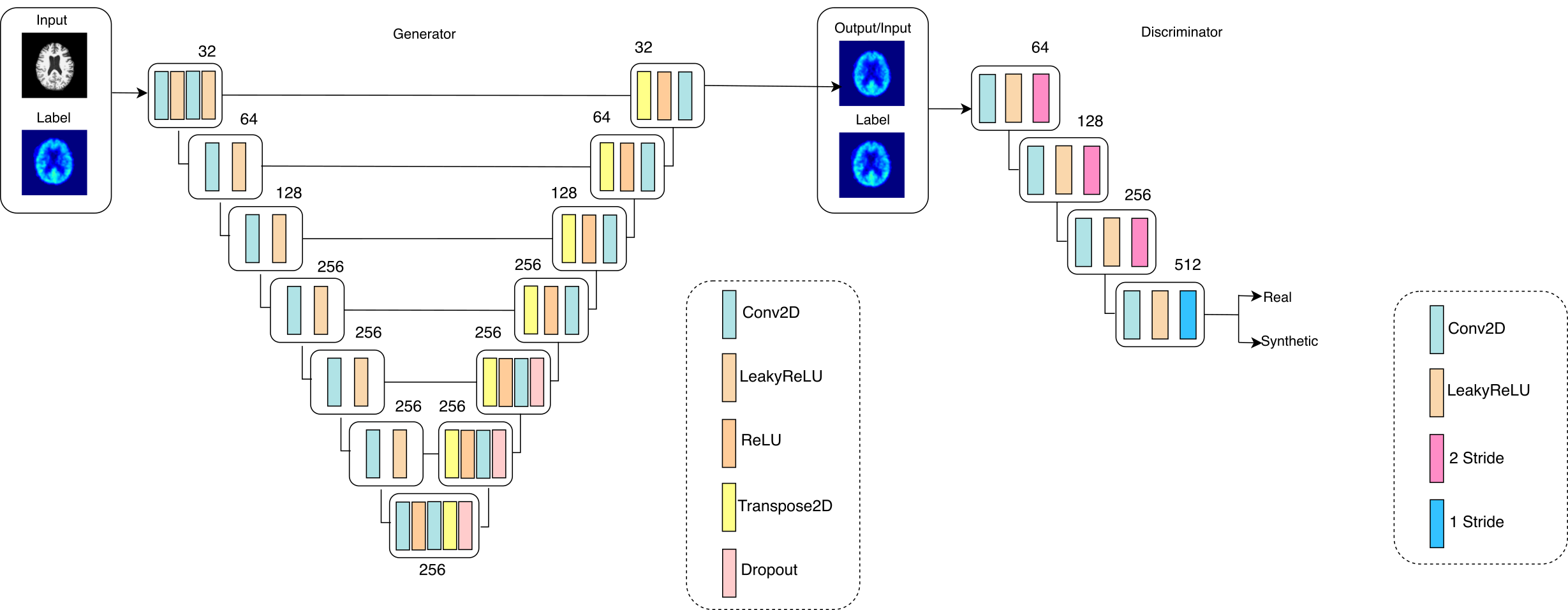}
\caption{Proposed image translation model following a cGAN architecture, normalization layers were removed as they cause the model to disregard the SUVr quantification. The Generator follows an enconder-decoder architecture that receives the input (MRI) and label (PET), the label conditions the model to produce an approximation of the label based on the input. The Discriminator follows a classification architecture that is also conditioned and determines if the produced image is synthetic or real.} 
\label{model_arch}
\end{figure}

\noindent For preprocessing, the dynamic PET images were converted to static by summing the last 30 minutes of the tracer passage, then PET images were brain extracted to eliminate scattering outside the head generated by the radiotracer, the MRI images were preprocessed with Freesurfer that produced the brain extractions. Both PET and MRI images were co-registered to the Montreal Neurological Institute (MNI) 1mm template using a composite transform that comprised translation and affine registration using Advanced Normalization Tools (ANTs), and split them in the axial plane.  

\noindent The cGAN architecture shown in Figure \ref{model_arch} was implemented to conduct image translation using the PyTorch deep learning library \cite{torch}.   A custom loss function was implemented to mask the image during training and only penalize information inside of the brain.  The model was trained with 550 subjects and stratified with equal proportion of females and males and different levels of cognitive impairment levels. 

\noindent The model was tested with 335 unseen MRI images to generate synthetic PET images that were evaluated with their paired true PET images by examining the quality of individual cases and the population with Structural Similarity Index (SSIM) \cite{SSIM} and Peak Signal-to-Noise Ratio (PSNR) \cite{PSNR}.

\begin{figure}[p]
\includegraphics[width=\textwidth]{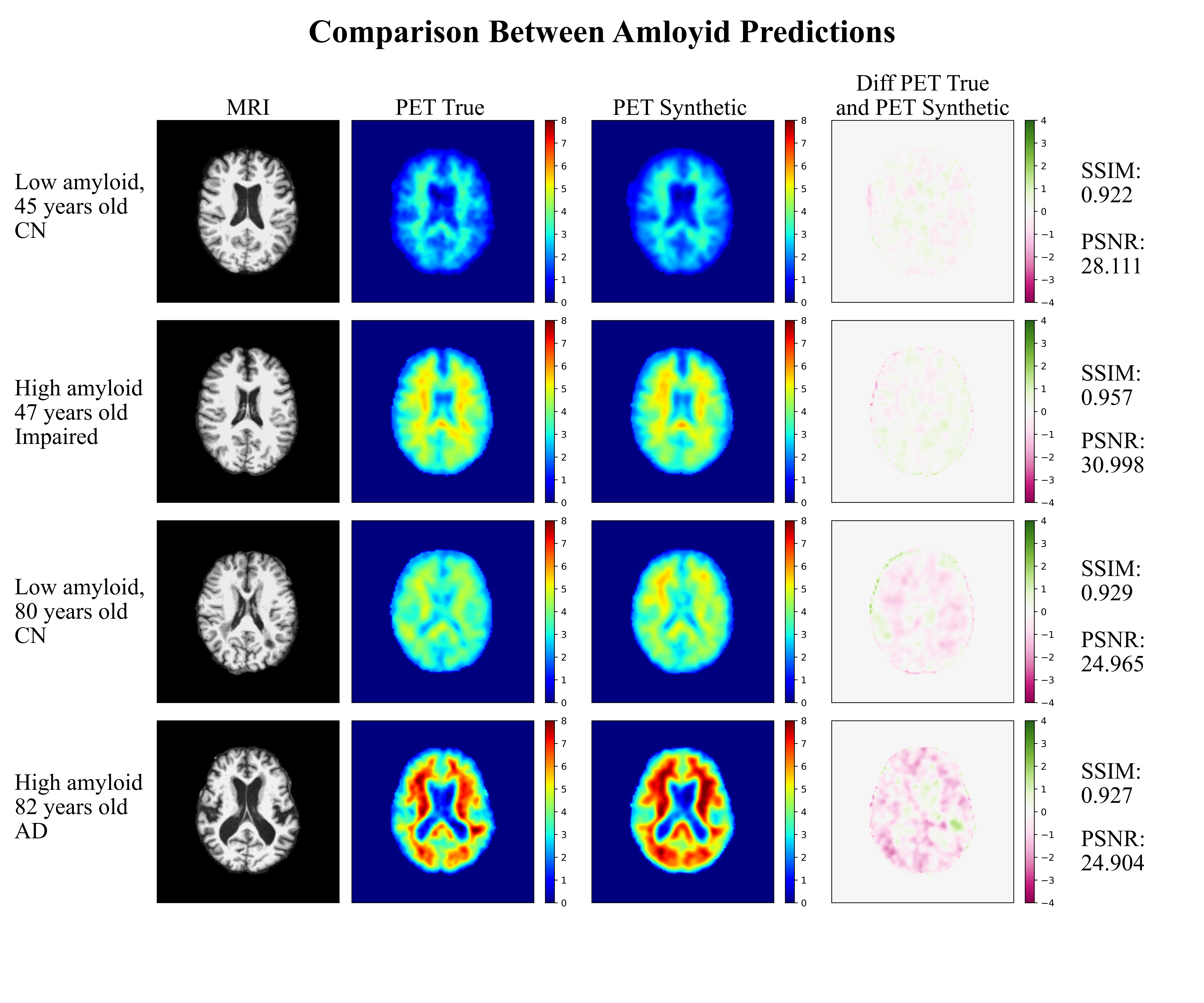}
\caption{Amyloid-beta predictions panel. Columns Left: MRI (input), center-left: PET True (label), center-right: PET synthetic generated by the model, right: Difference map between true and synthetic PET.   4 Subjects, 2 young where one is CN, the other is cognitively impaired and 2 old where one is CN and the other has AD. Based on the reported SSIM and PSNR the model can produce synthetic images that have high degree of similarity with the true in both shape and contrast.} 
\label{model_preds}
\end{figure}

\section*{Results}
\begin{figure}[p]
\includegraphics[width=\textwidth]{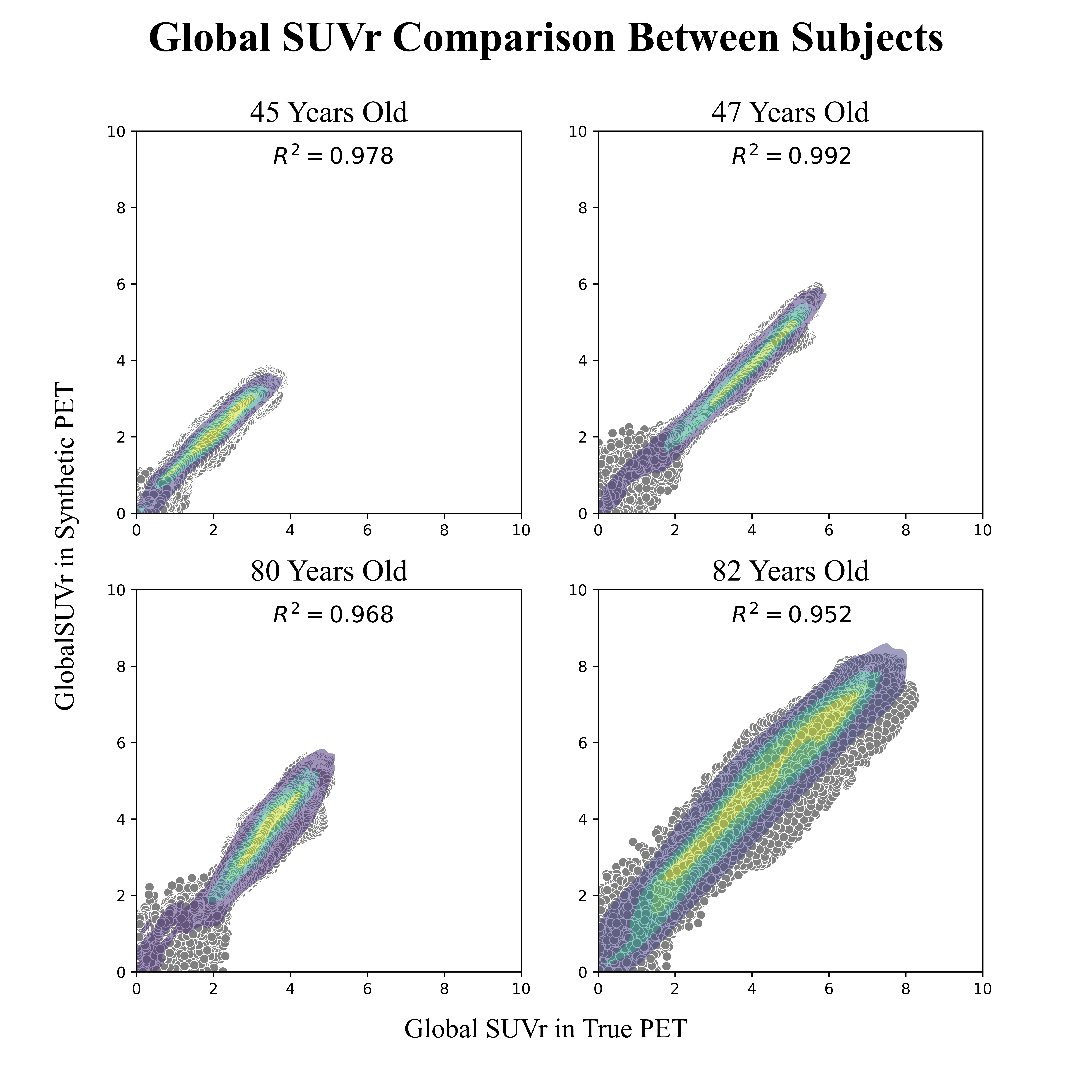}
\caption{Global SUVr comparison from the shown subjects in figure 2, the reported R2 above 0.95 indicates the model can estimate the quantification of the tracer based on the structural MRI alone. The SUVr density shows the model can generate quantification on the high levels of amyloid burden in the AD patient (80 years old) and impaired patient (47 years old).} 
\label{scatter_density}
\end{figure}

\begin{figure}[p]
\includegraphics[width=\textwidth]{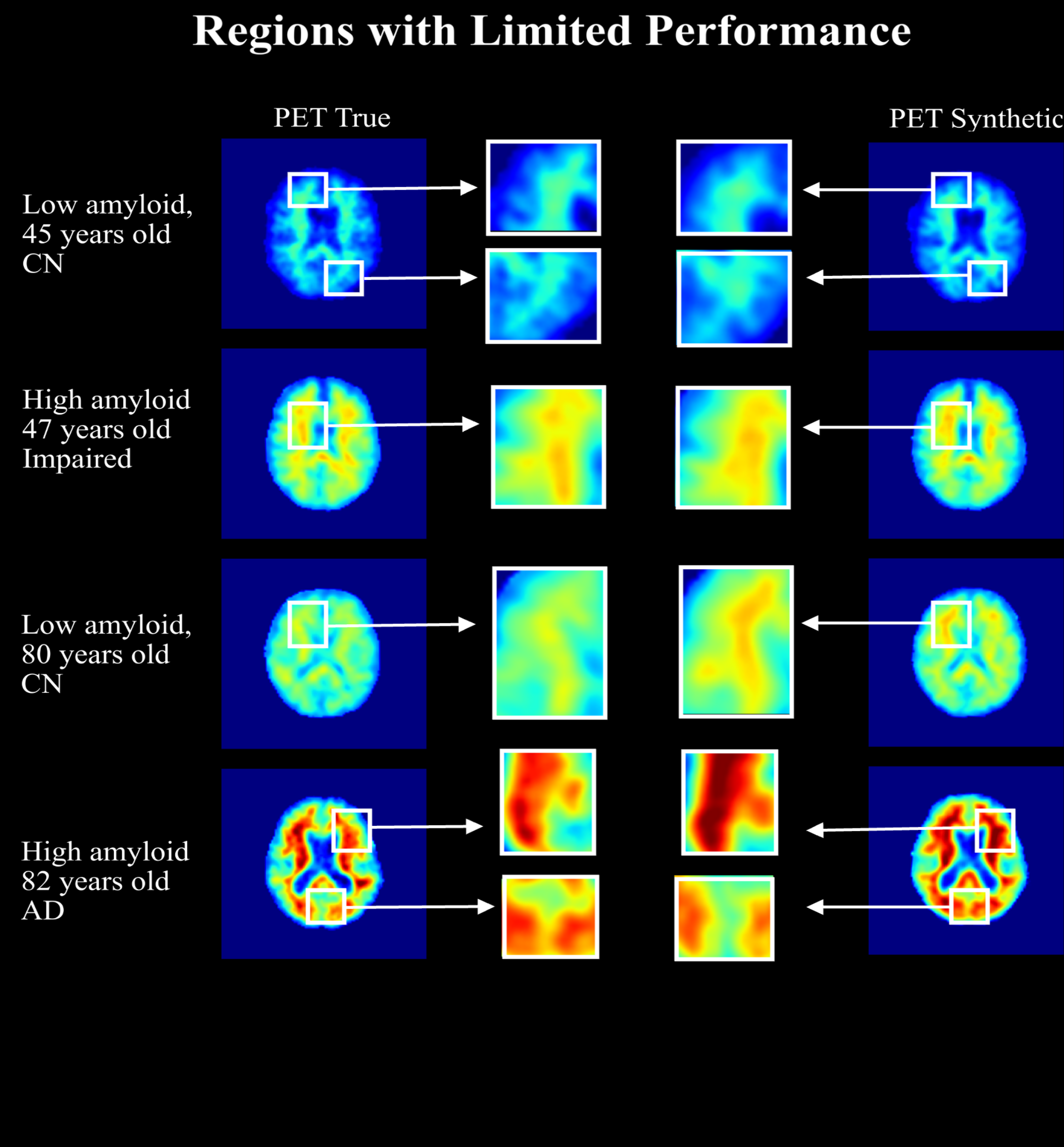}
\caption{Detailed visual comparison between predictions and true images, the image shows that there is still room for improvement in the quantification of SUVr for the images generated by the image translation model.  The cases that have low and mid-levels of amyloid show the quantification has high degree of accuracy, however, for the case of high amyloid the SUVr quantification still requires improvement.} 
\label{zoomin_panel}
\end{figure}

Four subjects were selected as examples showing how the model is able to produce high quality synthetic amyloid-beta PET images for both low and high amyloid cases, in both younger and older participants.  Figure \ref{model_preds} contains inputs, truth cases, synthesized images and difference maps.  The SUVR comparison in Figure \ref{scatter_density}, reports R2 > 0.95 for the four cases implying that the model is able to produce quantitative accurate synthetic amyloid-beta PET images.   Figure \ref{zoomin_panel} shows magnifications focusing on regions with limited performance.  

\begin{figure}[p]
\includegraphics[width=\textwidth]{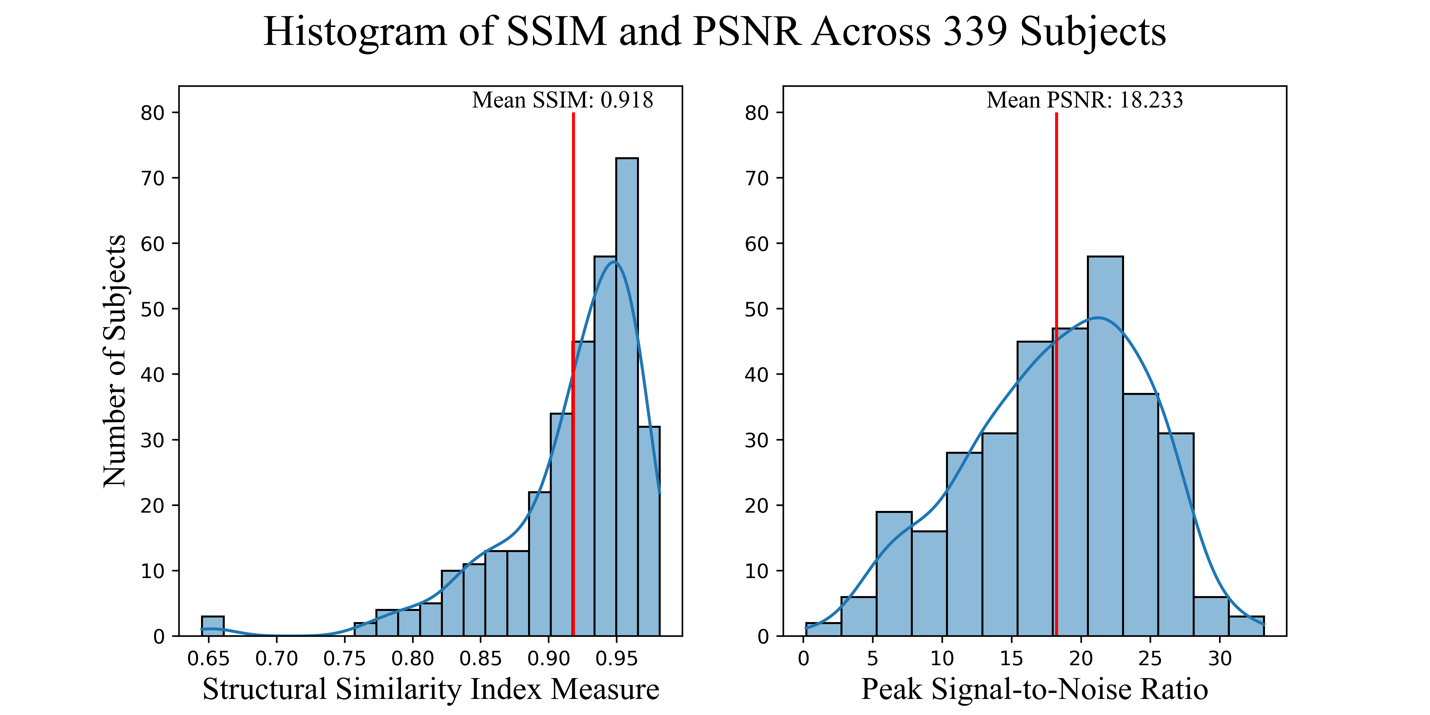}
\caption{SSIM and PSNR obtained from the 339 cohort of unseen subjects. A mean SSIM above 0.90 indicates that the model is able to produce images that have high degree of similarity. A high PSNR also indicates that the model created high quality synthetic PET images compared with the true PET. The SSIM histogram shows that most of the images fall between between 0.90-0.95, while the PSNR histogram shows the images have an overall high PSNR.} 
\label{Histograms}
\end{figure}

\noindent Figure \ref{Histograms} shows a high overall SSIM and PSNR meaning that the model generalizes well to the population of unseen images.

\section*{Discussion}
The future work for this project will include improving the performance, particularly of the small discrepancies seen in Figure \ref{zoomin_panel}, by exploring different ways of assessing the quantification during training.  This is a known problem in generative models as most of the implementations and metrics focus on generating relative images by normalizing the pixel values disregarding the quantification of the contrast \cite{GLA-GAN, BPGAN}.   Future work will also include developing a 3D model while targeting the quantification problem.  

\noindent Here-in, a pipeline to generate high-quality synthetic amyloid-beta PET images from structural MRI was implemented, the results are encouraging with inspected images having a high degree of similarity and population wide metrics show SSIM>0.9.  The method provides a non-invasive tool for early AD screening.

\section*{Acknowledgements:}
The authors would like to thank the University of Calgary, in particular the Schulich School of Engineering and Departments of Biomedical Engineering and Electrical \& Software Engineering; the Cumming School of Medicine and the Departments of Radiology and Clinical Neurosciences; as well as the Hotchkiss Brain Institute, Research Computing Services and the Digital Alliance of Canada for providing resources. The authors would like to thank the Open Access of Imaging Studies Team for making the data available.  FV – is funded in part through the Alberta Graduate Excellence Scholarship.  JA – is funded in part from a graduate scholarship from the Natural Sciences and Engineering Research Council Brain Create.  AE – was funded in part from the Biomedical Engineering Summer Studentship Program.  MEM acknowledges support from Start-up funding at UCalgary and a Natural Sciences and Engineering Research Council Discovery Grant (RGPIN-03552) and Early Career Researcher Supplement (DGECR-00124).  This work was made possible through a generous donation by Jim Gwynne.

\bibliographystyle{unsrt}
\bibliography{reference.bib}
\end{document}